\newcommand{\msbar}{{\overline {\rm MS}}}
\def\lsim{\raise0.2ex\hbox{$<$\kern-0.75em\raise-1.1ex\hbox{$\sim$}}}
\def\gsim{\raise0.3ex\hbox{$>$\kern-0.75em\raise-1.1ex\hbox{$\sim$}}}
\def\simgt{\rlap{\lower 4.0 pt\hbox{$\mathchar \sim$}}\raise 1.0pt \hbox {$>$}}
\def\simlt{\rlap{\lower 4.0 pt\hbox{$\mathchar \sim$}}\raise 1.0pt \hbox {$<$}}
\documentclass{PoS}

\title{SU(2) and SU(3) chiral perturbation theory analyses on meson and baryon masses in 2+1 flavor lattice QCD }

\ShortTitle{SU(2) and SU(3) ChPT analyses on meson and baryon masses in 2+1 flavor lattice QCD}

\author{PACS-CS Collaboration : 
\speaker{D. Kadoh}${}^{a}$\thanks{E-mail: kadoh@ccs.tsukuba.ac.jp},
 S.~Aoki${}^{b,c}$,
 N. Ishii${}^{a}$,
 K.-I.~Ishikawa${}^{d}$,
 N.~Ishizuka${}^{a,b}$,
 T. Izubuchi${}^{c,e}$,
 K.~Kanaya${}^{b}$,
 Y. Kuramashi${}^{a,b}$,
 Y. Namekawa${}^{a}$,
 M.~Okawa${}^{d}$,
 Y.~Taniguchi${}^{a,b}$,
 A.~Ukawa${}^{a,b}$,
 N. Ukita${}^{a}$,
 T.~Yoshi\'e${}^{a,b}$
 \\
 \llap{${}^a$}Center for Computational Sciences, University of Tsukuba, Tsukuba, Ibaraki 305-8577, Japan\\
 \llap{${}^b$}Graduate School of Pure and Applied Sciences, University of Tsukuba, Tsukuba, Ibaraki 305-8571, Japan\\
 \llap{${}^c$}Riken BNL Research Center, Brook-haven National Laboratory, Upton, New York 11973, USA\\
 \llap{${}^d$}Graduate School of Sciences, Hiroshima University, Higashi-Hiroshima, Hiroshima 739-8526, Japan\\
 \llap{${}^e$}Institute for Theoretical Physics, Kanazawa University, Kanazawa, Ishikawa 920-1192, Japan}

\abstract{We investigate the quark mass dependence of meson and baryon
masses obtained from
2+1 flavor dynamical quark simulations performed by 
the PACS-CS Collaboration. With the use of 
SU(2) and SU(3) chiral perturbation theories up to NLO, 
we examine the chiral behavior of the pseudoscalar meson masses 
and the decay constants in terms of the degenerate up-down quark mass 
ranging form 3 MeV to 24 MeV and two choices of the strange quark mass 
around the physical value. We discuss the convergence
of the SU(2) and SU(3) chiral expansions and 
present the results for the low energy constants.
We find that the SU(3) expansion is not convergent at NLO for the physical 
strange quark mass.
The chiral behavior of the nucleon mass is also discussed based 
on the SU(2) heavy baryon chiral perturbation theory up to NNLO.
Our results show that the expansion is well behaved only up to 
$m_\pi^2\approx 0.2$~GeV$^2$.}

\FullConference{The XXVI International Symposium on Lattice Field Theory \\
		 July 14 - 19, 2008\\
		 Williamsburg, Virginia, USA}

\begin{document}

\section{Introduction}
The PACS-CS Collaboration has been performing $N_F=2+1$ lattice QCD simulation 
with the nonperturbatively $O(a)$-improved Wilson quarks 
and the Iwasaki gauge action at the lattice spacing of $a=0.09$ fm on
a (2.9 fm$)^3$ 
box\cite{Aoki:2008sm,Y.Kuramasi:2008,N.Ukita:2008,K-I.Ishikawa:2008}. 
The DDHMC algorithm armored with several improvements
allows simulations in the light quark mass region, 
where the resulting pseudoscalar meson mass is down to
156 MeV. In this report we focus on chiral analyses of our data 
based on chiral perturbation theory (ChPT).  
We apply the SU(2) and SU(3) ChPTs up to NLO to the pseudoscalar meson
masses and the decay constants for the pion mass
ranging from 156 MeV to 411 MeV. 
The finite size effects are taken into account 
at one-loop level of the ChPTs.
We also examine  the quark mass dependence of the nucleon mass
employing the SU(2) heavy baryon ChPT up to NNLO.
The simulation details are given in separated reports\cite{Aoki:2008sm,N.Ukita:2008}.

\section{Simulation parameters}
We employ the $O(a)$-improved Wilson quark action with a nonperturbative
improvement coefficient $c_{\rm SW}=1.715$
and the Iwasaki gauge action at $\beta=1.9$, corresponding to the lattice spacing of $a=0.09$ fm, 
on a $32^3 \times 64$ lattice. Simulation parameters are summarized in
Table~\ref{table-1}, where the pion mass and the quark masses are given
at each combinations of $\kappa_{\rm ud}$ and $\kappa_{\rm s}$.
The quark masses are renormalized perturbatively in the $\msbar$ scheme
at the scale of $1/a$.
The statistics of each ensemble is given by the Molecular Dynamics (MD) time.

\begin{table}[h!]
\begin{center}
\caption{Summary of pion mass and quark masses perturbatively
 renormalized in the $\msbar$ scheme at the scale of $1/a$.}
\label{table-1}
\begin{tabular}{ccccccc}\hline
$\kappa_{\rm ud}$ & $\kappa_{\rm s}$  & $m_{\pi}$  & 
$m_{\rm ud}^{\msbar}$[MeV] & $m_{\rm s}^{\msbar}$[MeV] & MD time  \\
\hline
\hline
0.13781  &  0.13640  &  156     &   3.5    &   87  &   990 \\
0.13770  &  0.13640  &  296     &   12     &   90  &  2000 \\
0.13754  &  0.13640  &  411     &   24     &   92  &  2250 \\ 
0.13727  &  0.13640  &  570     &   45     &   97  &  2000 \\
0.13700  &  0.13640  &  702     &   67     &  103  &  2000 \\
\hline
0.13754  &  0.13660  &  385     &    21    &   77  &  2000 \\
\hline
\end{tabular}
\end{center}
\end{table} \vspace{-2mm}

\section{SU(2) and S(3) ChPT analyses for the pseudoscalar meson sector}
In principle our results should be compared with the predictions of the 
Wilson ChPT\cite{wchpt}.  However, their expressions in terms of 
the AWI quark masses agree with those of the continuum ChPT up to NLO 
by a redefinition of 
the low energy constants (LEC)\cite{Aoki:2008sm}.  We then 
discuss our results based on the continuum ChPT.

\subsection{SU(3) ChPT}
In the continuum SU(3) ChPT the one-loop expressions for 
$m_\pi$, $m_K$, $f_\pi$, $f_K$ contain six unknown 
LECs, $B_0, f_0, L_4, L_5, L_6, L_8$. 
We determine these parameters by applying
a simultaneous fit to 
$m_\pi/2m_{\rm ud},m_K^2/\{m_{\rm ud}+m_{\rm s}\},f_\pi,f_K$, where
the finite size corrections are taken into account 
at one-loop level\cite{Colangelo:2005gd}.

\begin{figure}[h!]
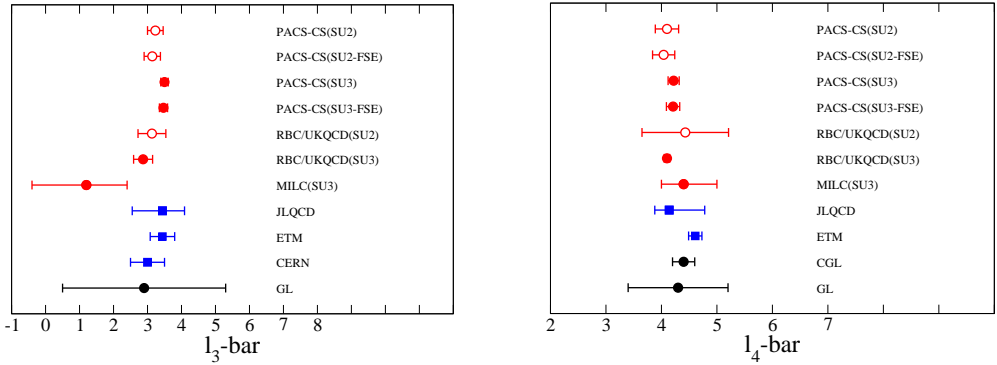

\begin{center}
\hspace{-0.6cm}
\includegraphics[width=6cm,keepaspectratio,clip]{lec_2f/l3-bar.eps}
\hspace{1cm}
\includegraphics[width=6cm,keepaspectratio,clip]{lec_2f/l4-bar.eps}
\caption{Comparison of the results for $\bar l_3$ and $\bar l_4$. Black symbols denote the phenomenological
estimates\cite{gasser,colangelo01}. Blue ones are for 2 flavor results\cite{del07,urbach07,jlqcd08}. Red closed(open) symbols represent the results for the SU(3) (SU(2)) ChPT
analyses in 2+1 flavor dynamical simulations\cite{Bernard:2007ps,Allton:2008pn}.  }
\label{fig:lec_2f}
\end{center}
\end{figure}

The results for the LECs are listed in
Table~\ref{table-2} where the phenomenological 
estimates with experimental inputs\cite{Amoros:2001cp}, 
the RBC/UKQCD results\cite{Allton:2008pn} and
the MILC results\cite{Bernard:2007ps} are also presented for comparison. 
We spot some discrepancies for the central values for the results of $L_{4,5,6,8}$.
Large errors make it difficult to draw any conclusions. 
We instead make a comparison in terms of the SU(2) LECs obtained
by the conversion from the SU(3) ones.  This is shown 
in Fig.~\ref{fig:lec_2f}. 
Our results are ${\bar l}_3=3.47(11)$, ${\bar l}_4=4.21(11)$ 
and ${\bar l}_3=3.50(11)$, ${\bar l}_4=4.22(10)$ with and without
the finite size corrections, respectively. Except for the MILC result 
for ${\bar l}_3$, all results are reasonably consistent.

\begin{table}[h!]
\begin{center}
\caption{Results for the LECs in comparison with the phenomenological
estimates, the RBC/UKQCD results and the MILC results.}
\label{table-2}
\begin{tabular}{c|cccccc}
            & w/o FSE    & w/ FSE      & phenomenology 
  & RBC/UKQCD 
  & MILC  \\
\hline
$L_4$ 	& -0.04 (10) &  -0.06(10) & 0.0 (0.8)    
  & 0.139 (80)  
&  0.1 (3) $(^{+3}_{-1})$ \\
$L_5$      	& 1.43 (7)	 & 1.45 (7)   & 1.46(10)       
& 0.872 (99) 
 &  1.4 (2) $(^{+2}_{-1})$ \\
$2L_6-L_4$ 	& 0.10 (2)   & 0.10 (2)   & 0.0 (1.0)  
    & -0.001 (42) 
&  0.3 (1) $(^{+2}_{-3})$ \\
$2L_8-L_5$  & -0.21 (3)  & -0.21 (3)  & 0.54 (43)    
  & 0.243 (45)  
&  0.3 (1) (1) \\
\hline
$\chi^2$/dof & 4.2(2.7) & 4.4(2.8) & $-$ & 0.7 & $-$ \\
\hline
\end{tabular}
\end{center}
\end{table}

\vspace{-5mm}

\begin{figure}[h!]
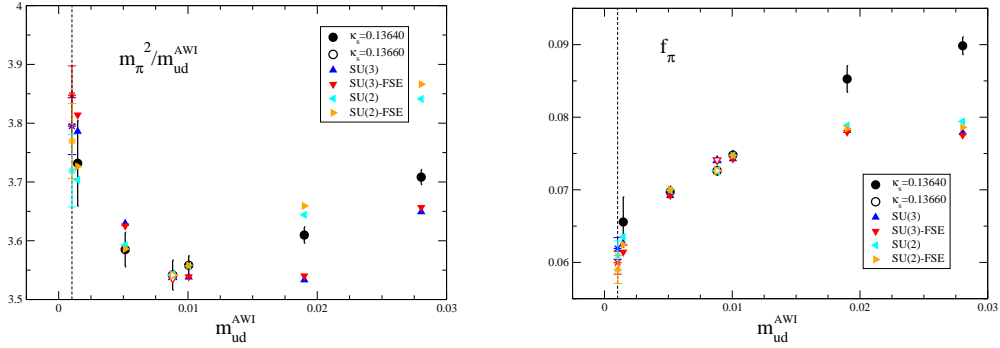

\begin{center}
\includegraphics[width=6cm,keepaspectratio,clip]{chfit_nf2nf3/mpi2.eps}
\hspace{1cm}
\includegraphics[width=6cm,keepaspectratio,clip]{chfit_nf2nf3/fpi.eps}
\caption{Fit results for $(am_\pi)^2/(am_{\rm ud}^{\rm AWI})$
 (left) and $f_\pi$ (right). The black  symbols 
represent the lattice results. 
The red and blue triangles denote the SU(3) fit results plotted
at the measured quark masses. 
The cyan and orange ones are for the SU(2) case. 
The open and filled symbols distinguish the results 
at $\kappa_{\rm s}=0.13640$ and 0.13660. 
The star symbols represent the extrapolated values at the physical point
 denoted by the vertical dotted line. }
\label{fig:chfit}
\end{center}
\end{figure}

Although we find that the SU(3) ChPT fit gives reasonable values for the
LECs, the value of $\chi^2$/dof is unacceptably large.
In Fig.\ref{fig:chfit} we plot the fit results for $m_\pi^2/m_{\rm ud}$ 
and $f_\pi$.  
The strange quark mass dependence between the data at 
$(\kappa_{\rm ud},\kappa_{\rm s})=(0.13754,0.13640)$ and $(0.13754,0.13660)$
is not well described. 
This flaw is the main cause for the large $\chi^2/$dof.

\begin{figure}[h!]
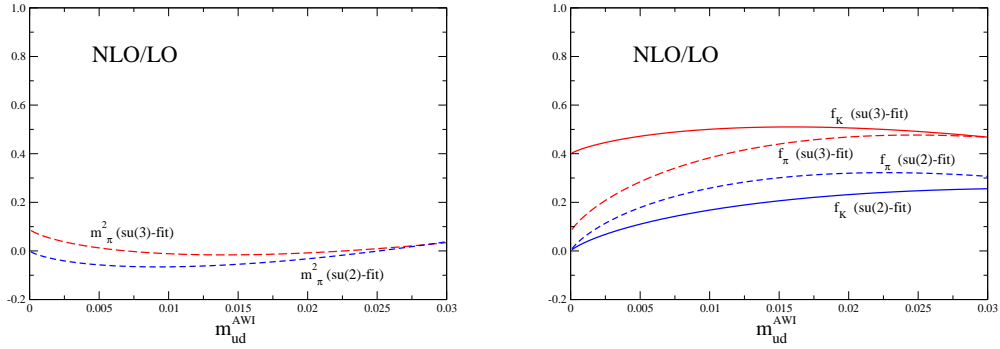

\begin{center}
\includegraphics[width=6cm,keepaspectratio,clip]{chpt_nlo2lo/ratio.mps.eps}
\hspace{1cm}
\includegraphics[width=6cm,keepaspectratio,clip]{chpt_nlo2lo/ratio.fps.eps}
\caption{Ratio of the NLO contribution to the LO one in the ChPT
 expansions for the pion mass and the pseudoscalar meson decay constants.
The strange quark mass is fixed at the physical value. }
\label{fig:chpt_nlo2lo}
\end{center}
\end{figure}

In order to investigate the origin of the discrepancy, we compare the NLO 
contributions with the LO ones in Fig.~\ref{fig:chpt_nlo2lo}. 
While the NLO is at most 10\% of the LO for the pion mass, 
the ratio is much larger for the decay constants, {\it e.g., }
for $f_\pi$ it
rapidly increases from 10\% at $m_{\rm ud}=0$  to  30\% 
around $m_{\rm ud}=0.01$. The situation
is worse for $f_K$: The NLO contribution is about 40\% of the LO one 
even at $m_{\rm ud}=0$,
most of which stems from the loops containing the strange quark. 
These observations mean that 
the physical strange quark mass is not small enough 
to be well controlled up to NLO in the SU(3) ChPT. 

\subsection{SU(2) ChPT}
\label{sec:su2}
Instead of extending the SU(3) chiral expansions 
from NLO to NNLO, we employ the SU(2) ChPT up to NLO
for further chiral analyses without increasing the data points.
Our strange quark mass is close enough to the physical value 
to allow us an analytic 
expansion of the SU(2) LECs around the physical strange quark mass.

\begin{table}[h!]
\caption{Cutoff, quark masses and pseudoscalar meson decay
 constants determined with $m_\pi,m_K,m_\Omega$ as physical 
inputs. Quark masses are renormalized in the $\msbar$ scheme
at the scale of 2 GeV.}
\label{table:physpoint}
\footnotesize
\hspace{0.5cm}
\begin{tabular}{ccccc}\hline
             & w/o FSE      & w/ FSE      \\
\hline
$a^{-1}$[GeV]      & 2.176 (31)  & 2.176 (31)   \\
$m_{\rm ud}^\msbar$[MeV] & 2.509 (46)  & 2.527 (47)   \\
$m_{\rm s}^\msbar$[MeV]  & 72.74 (78)  & 72.72 (78)      \\
$m_{\rm s}/m_{\rm ud}$ & 29.0 (4)   & 28.8 (4)  \\
\hline
\end{tabular}
\hspace{0.3cm}
\begin{tabular}{ccccc}\hline
             & w/o FSE      & w/ FSE      & experiment   \\
\hline
$f_\pi$      & 132.6 (4.5)  & 134.0 (4.2)  & 130.7 $\pm$ 0.1 $\pm$ 0.36  \\
$f_K$        & 159.2 (3.2)  & 159.4 (3.1)  & 159.8 $\pm$ 1.4 $\pm$ 0.44  \\
$f_K/f_\pi$  & 1.201 (22)   & 1.189 (20)   & 1.223 (12)   \\
\\
\hline
\end{tabular}
\end{table}

For $m_\pi$ and $f_\pi$ we employ the SU(2) ChPT formulae where
the low energy constants $B$ and $f$ are linearly expanded in terms of 
the strange quark mass:
$B=B_s^{(0)}+m_{\rm s}B_s^{(1)}$ and $f=f_s^{(0)}+m_{\rm s} f_s^{(1)}$. 
The K meson is treated as a matter field in the isospin 1/2 linear 
representation of the SU(2) chiral transformation, which is coupled to the 
pions in an SU(2) invariant way.
This assumption leads to the following fit formulae for $m_K$ and $f_K$:
\begin{eqnarray}
 m_K^2=\bar m_K^2+\beta_m m_{\rm ud}, \qquad
 f_K=\bar f \left\{1+\beta_f m_{\rm ud} 
- \frac{3}{4} \frac{2 B m_{\rm ud}}{16\pi^2 f^2} {\rm ln} \left(\frac{2B m_{\rm ud}}{\mu^2} \right)\right\},
\end{eqnarray}
where $\bar m_K^2$ and $\bar f$ are also linearly expanded 
in terms of the strange quark mass: $\bar m_K^2=\alpha_m+\gamma_m m_{\rm
s}$ and 
$\bar f=\bar f_s^{(0)}+ m_{\rm s} \bar f_s^{(1)}$.

We apply a simultaneous fit to $m^2_\pi/2m_{\rm ud},f_\pi$ and $f_K$ 
and an independent fit to $m_K^2$.
Evaluating the finite size effects based on the NLO formulae of 
the SU(2) ChPT for $m_\pi$ and $f_\pi$, we obtain
\begin{eqnarray}
&& m^{\rm ph}_{\rm ud}B=0.009345 (27) {\rm GeV}^2, \ f=124.8 (51) {\rm MeV}, \ \bar l_3=3.23 (21), \ \bar l_4=4.10 (20),
\end{eqnarray}
without taking into account finite size effects, and 
\begin{eqnarray}
&& m^{\rm ph}_{\rm ud}B=0.009332 (26) {\rm GeV}^2, \ f=126.4 (47) {\rm MeV}, \ \bar l_3=3.14 (23), \ \bar l_4=4.09 (19),
\end{eqnarray}
including finite size effects, 
where $m^{\rm ph}_{\rm ud}$ is the up-down quark mass 
extrapolated to the physical point. 
The results for ${\bar l}_3$ and ${\bar l}_4$  are plotted 
in Fig.~\ref{fig:lec_2f} for comparison. 
All the lattice results reside in the range
$3.0 \simlt l_3 \simlt 3.5$  and $4.0 \simlt l_4 \simlt 4.5$ 
except the MILC result for ${\bar l}_3$.

The fit results for $m_\pi^2/2m_{\rm ud}$ 
and $f_\pi$ are plotted in Fig.\ref{fig:chfit}.
In the SU(2) case we do not observe any discrepancy 
around $am_{\rm ud}=0.01$.
The resulting $\chi^2$/dof are 0.43(77) and 0.33(68) with and without the 
finite size corrections, respectively. 
These numbers are an order of magnitude smaller than the SU(3) case. 
In Fig.\ref{fig:chpt_nlo2lo} we illustrate the relative magnitude 
of the NLO contribution to the
LO one in comparison with the SU(3) case.
The convergences of the SU(2) chiral expansions for $m_\pi$ and $f_\pi$ 
are clearly better than the SU(3) case.

\subsection{Results for physical quark masses and pseudoscalar decay constants}

The up-down and the strange quark masses and the lattice cutoff
are determined with the choice of $m_\pi,m_K,m_\Omega$ as physical inputs. 
We employ the SU(2) ChPT for the chiral analysis on
$m_\pi$ and $m_K$ as discussed in the above subsection.
For the $\Omega$ baryon we use a simple linear fit formula 
$m=\alpha+\beta m_{\rm ud}+\gamma m_{\rm s}$.

In Table~\ref{table:physpoint} we summarize the results for the quark
masses, the lattice cutoff and the pseudoscalar meson decay 
constants together with the experimental values. 
Our results for the  quark masses are smaller than 
the estimates obtained by recent 2+1 flavor lattice 
QCD simulations\cite{Bernard:2007ps,Allton:2008pn}. 
We note, however, that we employ the perturbative
renormalization factors at one-loop level 
which should contain an uncertainty. For the decay constants 
we observe a good consistency 
between our results and the experimental values
within the error of $2-3$\%. Here also one-loop perturbative 
results for the renormalization factors are used.  We note that work is 
in progress to determine the renormalization factors non-perturbatively 
using the Shcr\"odinger functional~\cite{taniguchi}.

\section{SU(2) ChPT analysis for nucleon mass}
For the nucleon mass we employ the SU(2) heavy baryon ChPT 
formula upto NNLO\cite{Gasser:1987rb}:
\begin{eqnarray}
&&m_N=m_0-4c_1 m_\pi^2 -\frac{6g_A^2}{32\pi f_\pi^2}m_\pi^3 \nonumber \\
&&\hspace{1cm} +\left[e_1(\mu)-\frac{6}{64\pi^2f_\pi^2}\left(\frac{g_A^2}{m_0}-\frac{c_2}{2}\right)
-\frac{6}{32\pi^2f_\pi^2}\left(\frac{g_A^2}{m_0}-8c_1+c_2+4c_3\right) {\rm ln}{\frac{m_\pi}{\mu}}
\right]m_\pi^4 \nonumber \\
&&\hspace{1cm} +\frac{6g_A^2}{256 \pi f_\pi^2 m_0^2}m_\pi^5 + O(m_\pi^6),
\label{eq:nucleon_chpt}
\end{eqnarray}
which contains six new low energy constants $m_0,c_1,c_2,c_3,g_A, e_1$. 
The $O(m_\pi^5)$ term which is obtained by the 
relativistic baryon ChPT\cite{Procura:2003ig}
gives only a small contribution so that the results are little affected in the
following analyses.

\begin{figure}[h!]
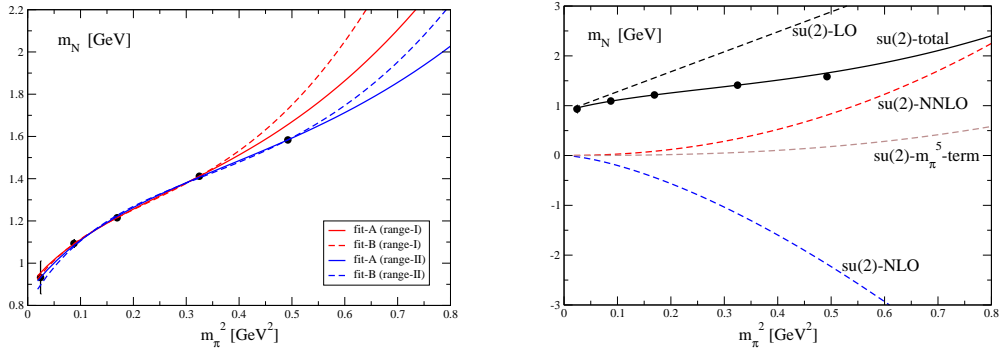

\begin{center}
\hspace{-0.6cm}
\includegraphics[width=6cm,keepaspectratio,clip]{nucleon_nf2/mpi2.mN.eps}
\hspace{1cm}
\includegraphics[width=6cm,keepaspectratio,clip]{nucleon_nf2/ratio.mN.eps}
\caption{Fit results of the nucleon mass using the SU(2) heavy baryon
ChPT formula
(left) and each contribution of the
 LO, NLO, NNLO and $O(m_\pi^5)$ terms in the case of fit-A with range-I 
(right).}
\label{fig:nucleon_terms}
\end{center}
\end{figure}

Since the number of our lattice data points is not sufficient for 
a full determination of the LECs by the ChPT fit,
we choose to fit $m_0,c_1,e_1$ while $g_A,f_\pi,c_1,c_2,c_3$ are 
fixed at the phenomenologically viable values.  
Following Ref.~\cite{Ali Khan:2003cu}
we set $g_A=1.267$, $c_2=3.2{\rm GeV}^{-1}$
and consider two possibilities for $c_3$: $c_3=-3.4 {\rm GeV}^{-1}$ 
as fit-A and $c_3=-4.7 {\rm GeV}^{-1}$ as fit-B.
For $f_\pi$ we use the pion decay constant 
extrapolated at the chiral limit with the SU(2) ChPT fit. 
We take two fit ranges for the data series with $\kappa_{\rm s}=0.13640$: 
Range-I is for $0.13781 \le \kappa_{\rm ud}\le 0.13727$ 
and range-II for $0.13781 \le \kappa_{\rm ud}\le0.13700$.
The fit results are depicted in Fig.~\ref{fig:nucleon_terms}.
We find that the lattice results are remarkably 
well described up to $m_\pi^2=0.5$ ${\rm GeV}^2$. 
Table~\ref{table:nucleon-lec} summarizes the results for the LECs 
together with the QCDSF/UKQCD results for comparison. 
The nucleon sigma term is also obtained 
through $\sigma_{N\pi}=m_\pi^2(\partial m_N / \partial m^2_\pi) $.
All the results are compatible within the $2\sigma$ errors.

In Fig.~\ref{fig:nucleon_terms} we separately draw the contribution of the LO, NLO,
NNLO and $O(m_\pi^5)$ terms in the chiral expansion.  
Observe that there is a drastic cancellation between the LO and NLO contributions 
which are both large in magnitude, and that the NNLO contribution monotonically 
increases as $m_\pi$ becomes heavier. 
The convergence of chiral expansion
in the SU(2) heavy baryon ChPT is hardly controlled 
beyond $m_\pi^2\sim 0.2$ ${\rm GeV}^2$. 



\begin{table}[h!]
\footnotesize
\hspace{5mm}
\begin{tabular}{|c|cc|cc|}\hline
           &  \multicolumn{2}{c|} {range-I} & \multicolumn{2}{c|}  {range-II}  \\
                           & fit-A & fit-B & fit-A & fit-B \\
\hline
\hline
$m_0$ [GeV]                      & 0.880 (50)   & 0.855 (47)   & 0.850 (27) & 0.795 (27)  \\
$c_1$  [GeV$^{-1}$]             & $-$1.00 (10)  & $-$1.19 (10)   &  $-$1.08 (4) & $-$1.34 (4) \\
$e_1(1{\rm GeV})$ [GeV$^{-3}$]      & 3.7 (1.4)  & 4.2 (1.4)         &  2.9 (4)  & 2.4 (3)  \\  
\hline
$\chi^2$/dof                     & 0.1 (0.9) & 0.0 (0.5)        & 0.3 (0.9)  & 1.2 (1.6)   \\ 
\hline
$\sigma_{N\pi} $ [MeV]       & 51.4 (7.6)   & 60.1 ( 7.3)   &  56.8 (2.7)  & 70.8 (2.6)  \\
\hline
  \end{tabular}
  \hspace{3mm}
  \begin{tabular}{|cc|}\hline
            \multicolumn{2}{|c|} {QCDSF-UKQCD}  \\
                    fit-A & fit-B  \\
\hline
\hline
 0.89 (6)   & 0.76 (6)     \\
 $-$0.93 (5)  & $-$1.25 (5)    \\
 2.8 (0.4)  & 1.7 (0.5)          \\  
\hline
\multicolumn{2}{c}{} \\
\multicolumn{2}{c}{}
\end{tabular}
  
 \caption{Results for $m_0,c_1,e_1$ together with the QCDSF/UKQCD results.
} 
 
\label{table:nucleon-lec}
\end{table}

 \vspace{-5mm}

\section*{Acknowledgment}
Numerical calculations for the present work have been carried out
on the PACS-CS computer
under the ``Interdisciplinary Computational Science Program'' of
Center for Computational Sciences, University of Tsukuba.
This work is supported in part by Grants-in-Aid for Scientific
Research from the Ministry of Education, Culture, Sports, Science and
Technology (Nos.
16740147,   
17340066,   
18104005,   
18540250,   
18740130,   
19740134,   
20340047,   
20540248,   
20740123,   
20740139    
).


\begin{thebibliography}{99}

\bibitem{Aoki:2008sm}
  PACS-CS Collaboration, S.~Aoki {\it et al.},
  arXiv:0807.1661 [hep-lat].
  
 \bibitem{Y.Kuramasi:2008} 
 Y.~Kuramashi for PACS-CS Collaboration, these proceedings.
 
 \bibitem{N.Ukita:2008}
 PACS-CS Collaboration, N.~Ukita {\it et al.}, these proceedings.

  \bibitem{K-I.Ishikawa:2008} 
 K.-I.~Ishikawa, these proceedings.

\bibitem{wchpt}
S.~Aoki, O.~B{\"a}r, T.~Ishikawa and S.~Takeda,
Phys. Rev. {\bf D73}, 014511 (2006); S.~Takeda,
{\it Wilson chiral perturbation theory for 2+1 flavors},
Doctor Thesis (unpublised).


\bibitem{Colangelo:2005gd}
  G.~Colangelo, S.~D{\"u}rr and C.~Haefeli,
  Nucl.\ Phys.\  B {\bf 721} (2005) 136.



\bibitem{Amoros:2001cp}
  G.~Amor{\'o}s, J.~Bijnens and P.~Talavera,
  Nucl.\ Phys.\  B {\bf 602}, 87 (2001).


\bibitem{Allton:2008pn}
  C.~Allton {\it et al.},
  arXiv:0804.0473 [hep-lat].



\bibitem{Bernard:2007ps}
  C.~Bernard {\it et al.},
  PoS {\bf LAT2007}, 090 (2007).



 
\bibitem{gasser} 
J.~Gasser and H.~Leutwyler,  Ann of Phys. {\bf 158} (1984) 142;  
Nucl. Phys. {\bf B250} (1985) 465.

\bibitem{colangelo01}
G.~Colangelo, J.~Gasser and H.~Leutwyler,
Nucl. Phys. {\bf B603}, 125 (2001).

\bibitem{del07}
L. Del Debbio {\it et al.}, JHEP {\bf 0702}, 056 (2007);
JHEP {\bf 0702}, 082 (2007).

\bibitem{urbach07}
C.~Urbach, PoS {\bf LATTICE2007}, 022 (2007).

\bibitem{jlqcd08}
JLQCD Collaborations, J.~Noaki {\it et al.}, arXiv:0806.0894[hep-lat].

\bibitem{taniguchi}
Y. Taniguchi for PACS-CS Collaboration, these proceedings.

\bibitem{Gasser:1987rb}
  J.~Gasser M.~E.~Sainio and A.~Svarc,
  Nucl.\ Phys.\  B {\bf 307}, 779 (1988).

\bibitem{Procura:2003ig}
  M.~Procura, T.~R.~Hemmert and W.~Weise,
  Phys.\ Rev.\  D {\bf 69}, 034505 (2004)
  [arXiv:hep-lat/0309020].


\bibitem{Ali Khan:2003cu}
  QCDSF-UKQCD Collaboration,
  A.~Ali Khan {\it et al.},
  Nucl.\ Phys.\  B {\bf 689}, 175 (2004)
  [arXiv:hep-lat/0312030].


  
\end{thebibliography}
\end{document}